# Towards a secured smart IoT using light weight blockchain: An aim to secure Pharmacy Products

**Md. Faruk Abdullah Al Sohan**
Department of Computer Science, American International University-Bangladesh, Dhaka
Email: faruk.sohan@aiub.edu

**Samiur Rahman Khan**
Department of Computer Science, American International University-Bangladesh, Dhaka
Email : samiurk70@gmail.com

**Nusrat Jahan Anannya**
Department of Computer Science, American International University-Bangladesh, Dhaka
Email : nusratanajahan786@gmail.com

**Dr. Md Taimur Ahad**
Department of Computer Science, American International University-Bangladesh, Dhaka
Email : taimur.ahad@aiub.edu

-----------------------------------------------------------------ABSTRACT-----------------------------------------------------------------

Blockchain has proven a very developed and secured technology. It ensures data integrity with authentic connected nodes. Now-a-days, blockchain with IoT is a great combination for secured and smart end to end product delivery. This observation has motivated the research to develop a conceptual model to provide a secure pharmaceutical product delivery by developing a IoT integrated with lightweight blockchain. The undeveloped and most of the developing countries are facing problems such as drug counterfeits, shortages, opiates and tracking them became difficult because of less transparency. Also, nature sensitive medicines need to be stored under controlled temperature known as cold-chain shipping. The storage of these information in the recent software is done in the centralized databases that is prone to data manipulations and hacks. Due to less production drugs needed to be imported with maintaining drug supply chain regulations by law. This paper proposes a lightweight blockchain model for pharmaceutical industries by using IoT. This model ensures traceability of drugs within a very simple way which is less complex compared to the existing ones.

## 1. INTRODUCTION

IoT, the Internet of Things is a system of interconnected devices with unique identification (UIDs) and a network communication capacity without needing contact between human beings or human beings. An IoT eco system consists of web smart devices, used for the collect, distribute and act on data that they collected from their environments, using integrated systems including processors, sensors, and communication hardware. IoT devices exchange sensor data that they collect through the link to an IoT portal or other edge system where they either send data to the cloud for analysis or local analysis. Often, they interact and act on the information they receive from each other with other similar devices [1]. Kevin Ashton stated this concept in 1999 [2]. IoT based devices are wirelessly linked with each other. It is estimated that by 2025, there will be 75 billion IoT devices, and an article claims that IoT security problems are mostly ignored [3].

Blockchain (BC) is a technology invented by Satoshi Nakamoto in 2008 [4]. The basic goal of the BC is to store data and information digitally without the intrusion of a third party or a centralized system [5]. On the contrary of Centralized network, Decentralized network system has gained quite a monetary and distinguishable value on the global IT sector of the world. Staging itself a brand-new economy based on cryptocurrency where the estimated monetary value outreaches 1$ trillion [6]. The technological system which delivered such high scale stable market for cryptocurrency is the Decentralized Blockchain environment. One of the bright sides of such a technical environment is overcoming the

security Secure sharing, Anti-money laundering, Real-time IoT, Personal identity security shortcomings of the centralized network models. Moreover, such features of this technology can be implemented over the revolutionary IoT systems that can benefit in all the IoT impacted sectors. BC will become one of the upcoming technological revolutions owing to its key characteristics: no central authority, no middlemen, real-time settlement, substantial cost reductions, and high degrees of transparency. It may be used in a variety of fields, including governance, healthcare, banking, the Online of Things, energy internet information security, public and social services, reputation systems, and education. In [7], they provide secrecy by employing the AES algorithm before making transactions, as well as other features such as real-time online verification, third-party verification, usability, and revocation when issuing certificates for students. However, incorporating BC into IoT is a difficult endeavor since delivering the vast number of resources and power that BC needs is difficult, if not impossible, for IoT devices. The major goal is to create a system that is more secure and speedier while requiring no high-power energy usage [8]. In the pharmacy sector, blockchain can play an important role to make the existing system more flexible. The pharmacy supply chain management system itself is one of the core sectors of the economy and public welfare of any country where the production, supply management and distribution takes place which in the current and previous time being has faced many stability issues and obstacles that objectively put it in a position of revaluating its fundamentals which mainly addresses its work mechanisms. The complexity of the structure for such a mechanism generates both technical and real-time problems like data validation, transparency issues, information inaccuracy and source of data. Furthermore, drug counterfeiting has been a major problem this system still cannot address, and it brings the Healthcare industry to a financial loss like mislabeling drugs with and selling it as the legit one [9]. This counterfeiting is way worse in the third world countries where illegal manipulation of drugs and drug trafficking is a constant process. Therein the current system having no other way to tackle the aforesaid issues, requires a heavy progression such that could c Medicine being one of the fundamentals needs of the public relies solely on the pharmacy supply chain management system to deliver it to the consumers and in doing so ensure the validity of the drug or the pharmaceutical product. This supply chain functions in a complex web of self-sufficient entities that regulate resources, raw materials, production, receiving etc. and they are commonly termed as supplier, distributor, manufacturer, patients, pharmacies, and hospitals [10]. These functionalities of this network cannot be tracked due to lack of a centralized entity and relationships among the stakeholders which provides us with no information on few entities. As such counterfeit products can easily be produced in this system and mislabeled to match with the source product and will appear as genuine. One of the newer technologies that emerged and bloomed in the past decade and got widely accepted in the global IT community is none other than Blockchain. It features Smart Contracts that makes this technology a game changer in the business and enterprise system like the Supply Chain Management System (SCMS) [11]. The tracing mechanism for this technology has been research and development and was reported that it could be applied in vehicle tracing, academic certifications, IOT integration and pharmaceutical industry [12].

In this research, we developed a novel model that might improve system performance. It is structured as follows: Section II discusses related works. Figures are used to describe the proposed model in Section III. Section IV describes the mining process. The system's operation is explained in Section V. Section VI discusses traceability, while Section VII discusses security and Section VIII delves into the conclusion.

## 2. RELATED WORKS

Blockchain has developed as a ground-breaking technology that safely, fault-tolerant, and transparently holds and transmits information. Blockchain has demonstrated the ability to transform the conventional supply chain industry into a supply chain that is encrypted, digital, anonymous, continuous, readable, and decentralized. According to distributed ledger-based technology, this is possible. The blockchain has the full chance to generate it safer, effective, accessible, and decentralized for any organization.

Pharmaceutical companies have been struggling to monitor their drugs during the supply chain process since the last decade, empowering counterfeiters to sell their fake medicines. Such medications do not help patients recover from the illness, but they may have several other harmful side effects. Counterfeit products are evaluated internationally as a very massive challenge for the pharmaceutical industry. A new blockchain and machine learning-based drug supply chain management and recommendation framework have been proposed being introduced by Khizar Abbas, and Wang-Cheol Song. Through using this machine learning-based drug supply chain management system, each user of the system will control the delivery of the drug. In addition, between the linked peer nodes, the proposed system can execute CRUD (create, read, update, delete) operations. In health care, the scheme is using machine learning algorithms and blockchain technologies with excellent performance. This method allows pharmaceutical firms to reduce the issue of counterfeit drugs and to improve business dramatically [13].

With the rise of Internet pharmacies, it has become increasingly difficult to standardize medication protection. Because these pharmaceuticals go via many complicated dispersed networks, counterfeits are difficult to identify, offering possibilities for counterfeits to enter the actual supply chain. The safeguarding of the pharmaceutical supply chain is becoming a serious concern for human health, which is a collective operation. Faisal Jamil and Lei Hang have presented a novel medicine supply chain management system based on Hyperledger Fabric and blockchain technology to handle secure drug supply chain data. The proposed system is distinguished from existing blockchain-based systems by the fact that it is built on a licensed network. This feature allows only a valid participant to participate and enroll in the blockchain network through a user identity manager. The user identity manager oversees issuing user enrollment and authentication certificates. Their tests revealed that using blockchain technology enhances throughput and reduces latency in the proposed system while using less resources [14].

A blockchain-based approach has been introduced by Kavita Kumari and Kavita Saini that can monitor the movement of drugs across the supply chain, from the distributor to the end patients. This system will help in fighting the counterfeiting of drugs and reduce them to great extent. Data handling and Drug traceability are very important in blockchain technology in the healthcare [15].

Jen-Hung Tseng and Yen-Chih Liao propose gcoin blockchain to establish straightforward drug transaction data as the basis of the drug data flow. All products that are involved in the drug supply chain would be able to participate simultaneously to prevent counterfeit drugs and to protect public health, including patients [16].

Bidi Ying, Wen Sun et al. proposed a secure blockchain model based on the patient's prescriptions. In the model, there were different blocks with different duties. The blocks are Health Department Center (HDC), Pharmacy Agent (PA), Pharmacy Industry (PI). All the blocks are trustworthy. HDC is responsible for registration. They register the PAs. PAs are responsible for communicating with the users. It also provides a connection with the PI. PI provides sufficient medicine supply in the pharmacy. It can also trace the medicine from the beginning. Overall, the system provides a secure drug supply in the health-care System [17].

Rajani et al. focus on enhancing blockchain systems to make them appropriate for IoT-based supply chain management systems since blockchain and IoT sensors play an important role in supply chain management systems. To create a medicine supply chain, the researchers use blockchain and IoT. The Raft consensus mechanism was employed to improve the system's output, while the Blockchain Distributed Network (BDN) and bloXroute servers were deployed to increase network scalability. To maintain the system's security integrity, an innovative hybrid method combining numerous cryptographic approaches was applied. The system's and its components' needs may be altered, making the framework adaptable. Furthermore, by employing bloXroute to solve the network bottleneck issue, the suggested architecture may support a higher consensus method with greater fault tolerance. The researchers develop and install the system, as well as undertake numerous performance and security analyses on a blockchain network [18].

The paper researched the possible ways to use blockchain technology in pharmaceutical domain to detect forgery or poor-quality medicinal products. Also, the regulations regarding the system were also discussed. The cause for the legal service of the blockchain technology in pharmaceutical domain were examined to protect the rights of patients. It revealed the process of the application of the technology in pharmaceutical activities. It was acknowledged that the legal basis of using blockchain technology was not perfect. Using blockchain technology reduced the hazard of supplying forgery medicines to patients, ensured clarity of delivery and less organizing costs [19].

Ahamad Musamih, Khaled Salah and et al proposed an approach for drug traceability in healthcare supply chain. In the model they used smart contract. There were also stakeholders. The whole system works on the Application Program Interface (API) which is used decentralized storage system. It uses different types of APIs such as Web3, JSON RPC, Infura, and so on. The stakeholders collect data with the help of the smart contact. In the stakeholders, there are some agencies. Decentralized Storage System is very low in cost. Mainly the stakeholders are responsible for tracing the drugs. Some of the agencies are always tracing the drugs from the manufacturing process. The system provides a good result rather than the existing one [20].

In [21], the authors implement a model for the patients using cloud computing which is able to give fast relief treatment to prevent multiple times admitting to hospital. This application can be used any organization to store the necessary documents. It provides high security as here Attribute Based Encryption (ABE) is used. It is a unique and not easily hackable application.

Drugledger, a structure aiming blockchain system to trace and regulate drug was proposed by Huang et al. The following system recreated an entire service architecture by isolating service provider in three individual service tools such as Certificate service provider (CSP), query service provider (QSP), and anti-attack service provider (ASP). The traditional p2p architecture's solutions ensured flexibility. The paper suggested to cut back blockchain depend on the drug's date of expiration and provide the corresponding algorithm, which helps the blockchain to gain a stable storage that seemed to be disregarded in the blockchain academic community that cares more about absoluteness rather than usability and feasibility [22]. Drugledger can contradict sybil attack and assured the stakeholders' traceability information's privacy and authenticity without losing flexibility of the system. An end to the data storage assured scalability with time being in storage [23].

## 3. CORE COMPONENTS

This section describes the main components of the proposed model, as seen in Fig 1. The core components are Block-chain, Overly Network, & Cloud Storage.

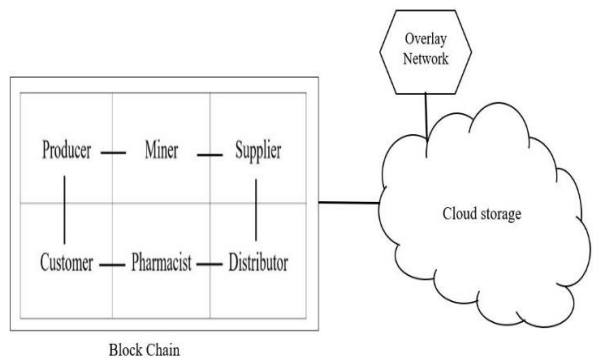

Fig 1. Overview & core components of the model

### 3.1. Blockchain

Block-chain is one of the core components of this model. It consists of six blocks. They are Producer, Miner, Supplier, Distributor, Pharmacist & Customer. All are connected to each other.

- **Producer:** Producer stores the raw elements of the drugs and by using them they produce drugs according to the demand. Prior to legal clearance, a pharmaceutical firm must do appropriate research and development work before it is permitted to produce a medicine. A medication is finally able to be commercialized and launched into the market after undergoing several producer and quality assurance procedures. The amount of product produced by the producer is determined by the needs of the consumer. Before a medicine may be put into production, it must first pass a series of tests and meet strict criteria. by maintain all the procedure, they produce drugs and sent them to the miner for further processing.

- **Miner:** Miner receives medicine from the producer. Miner plays a significant role for ensuring genuine drugs. Miner's job is to make sure that the quality of medicine is right. If the quality of medicine is maintained all the conditions, then Miner will deliver the medicines to the supplier and if not, then miner will return the medicines to the producer for reproduction. At the same time, miner will check the expiration date of medicine. As a result, the quality of the medicine is ensured before going to the supplier.

- **Supplier:** At any point of the life cycle of a product, suppliers play a significant role. It is essential for suppliers to guarantee the best quality and price for their distribution partners to preserve the faith among them. It will further ensure future repeat business. A new client will visit a supplier because they need anything in advance. We must have it ready for them. All is urgently required for those who require emergency medicine supplies. It is imperative that suppliers always have a diverse supply. Supplier sends the pharmaceutical drugs to the distributor. The supplier rechecks the expiration date of the medicine before handing it over to the distributor. If the expiration date of the medicine is sufficient then, the supplier sends the medicine to the distributor, otherwise the supplier returns the medicine to the producer.

- **Distributor**: Prescription drugs and other healthcare products are purchased directly from pharmacy suppliers and stored in pharmacies and distribution facilities throughout the country by pharmaceutical distributors. Drugstores, medical clinics, and service providers with state and county approvals place orders with distributors for the medications and supplies they need, and the distributor manages and provides the orders on a regular basis. Primary distributors will accept orders for ongoing operations to send them so quickly as possible to their destination. It always maintains the quantity of the drugs.

- **Pharmacist**: When the medicine is transferred to the pharmacy, a medication pharmacist is responsible for providing the customers with the right dosage and type of the medicine. The delivery of medicine is the last phase of the supply chain for pharmaceutics. Along with dispensing medications, this final phase includes after-sales services such as supplying accurate reports to consumers and processing refund requests, ensuring that customers get completely profit from the medication they bought. It also ensures that the drugs are genuine.

- **Customer**: A customer will visit a pharmacist to take the medicine as per the requirement. Pharmacist needs to ensure that customers get the best quality of medicine. Customers will also check the quality of the medicine to ensure that if he has received superior quality medicine. If he/she finds any fault, he/she can return the drugs to the producer directly.

## 3.2. Overlay Network

It establishes logical or physical connections between nodes in a network design. All the blocks in a blockchain technology create a peer-to-peer network, which refers to the distributed system that highlights the notion of the overlay network. It provides a network-like capability that allows you to operate on top of another network. Because of the overlay network, it is now easy to add or remove new blocks from the system.

## 3.3. Cloud Storage

Cloud storage is a popular and efficient data storage. It provides more data security with an extremely low cost. The blockchain technology deals with a huge data. Because of having huge data, sometimes it may not provide faster communications. All kinds of data will be stored in the cloud storage. In the cloud storage for this model, there are two components. One is for storing only list of medicines and another one is used for storing all the hashes and the transaction details that means data. All the blocks can access the cloud storage. It helps to reduce the pressure of the blockchain.

## 4. MINING PROCESS

The term "Mining Process" refers to the creation of new Blocks in the Blockchain ledger but one of the key functionalities of this process is also to verify all the transaction listings according to legitimate information [24]. In our conceptual model this process can help in evaluating the medicine listing using the right components and requirement of the clients. This process can be varied and modified according to the needs of a specific system. As we know that blockchain itself is an immutable ledger that helps in transaction recording and assets tracking in the given network, the transactions within these blocks are made verified and secured using the mining process [25,26]. It validates the information with the help of miners who reside in the same network as the blockchain, and they utilize a set of methodologies for the evaluation purpose [25]. Such techniques are followed by all the miners in the network and if there is any wrong data provided in the transaction it will be shared among all the miners so that every detail in each step of the process is shared for verifying the legitimacy of the list of transactions [26]. In case of our conceptual system model setting some necessary procedures of the miners to follow can help in validating all the medicine list with respect to quality, drug effectiveness and required amounts from the clients.

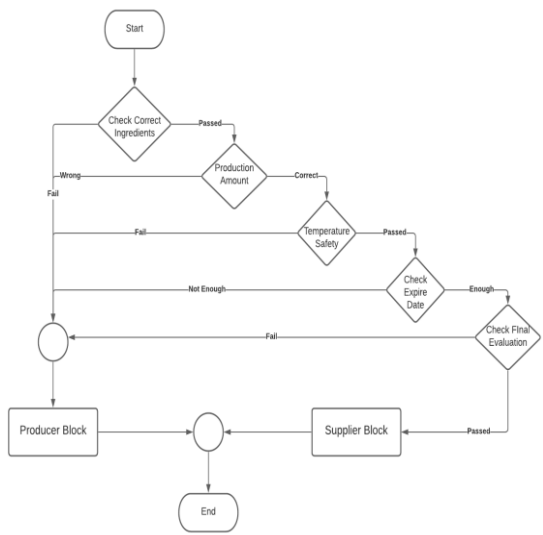

Fig 2. Mining Process of the system

From the Fig 2. we can observe how the overall mining process and the Evaluation Procedures followed for our conceptual model to ensure the best medicine supplied to the client.

### 4.1. Checking Correct Ingredient

In the first step the miners check the ingredient of the specific medicine from the medicine list, and they verify the authenticated amount of ingredients required in that medicine from the pharmacopeia and drug formulary provided in the

4miner network for this purpose. If the medicine in the list is found not to match with the correct amount of ingredients, then it will be sent to the Producer Block with a message regarding its inaccuracy in ingredients and if it is found to match correctly with the given directory it will move to the next stage of the evaluation procedure.

### 4.2. Production Amount

It is the second step where the miners evaluate the quantity of medicines requested from the client or distributor. Verification is mainly done by checking the demand of that medicine requested from the client and by monitoring the past orders of that medicine from hospitals. If the quantity is found to be not sufficient then it's sent to the producer block with message entitled as Insufficient quantity else the miners move on the next stage.

### 4.3. Temperature Safety

Medicines need to be preserved after manufacturing or its property may get damaged. Reports of the temperature conditions of all the medicine list can be accessible by the miners who then evaluate all the medicine list to confirm if they are well maintained or not. If report of any medicine list is missing or from the reports any of them are found to be preserved in wrong temperature, then it will be sent to the producer block reporting "Unsafe Temperature" as comment. The miners will move to the next phase with the correctly preserved medicine lists.

### 4.4. Expiry Date

In many cases expired medicines are sold off in many pharmacies and the expiry dates are found to modified which is very harmful for the customers. In our conceptual model the production takes place within the blockchain network so all the registries of the medicines produced are stored within our network which helps the miners to validate the authenticity of the expiry date and since blockchain is an immutable platform [26] the expiry dates can never be changed or adulterated. So, if any medicine is expired it will be sent to the producer block labelled "Date Expired" and miners will move on to the next stage with the rest of the medicine lists.

### 4.5. Quality Assurance

Here the miners determine the standards for safety testing of products from the quality assurance reports provided by the manufacturer of all the medicines in the list. Quality related issues found from quality control testing is reported to the miner network which helps in evaluating the quality assurance status of all the transactions in the medicine lists. If the requirements are not met, then it will be sent back to the producer block reporting quality assurance problem as a comment otherwise, it will be sent to the supplier block as all the steps in Evaluation Procedure are satisfied.

## 5. WORKING PRINCIPLE

The major goal of the system is to make people's lives simpler by providing a traceability feature for the drug. As blockchain is a very secure system, the users can give their personal information without any hesitation. By using this system, the users can trace their medicine from the beginning. The system also provides enough drugs, as a result, there will not arise any shortage or excess of drugs. As all the bocks of the blockchain can trace the drugs so, it ensures that the drugs are genuine. Without any hesitation, anyone can take these drugs. The system defines a set of tasks that must be followed for drugs to be delivered. The model's operation is depicted in Fig 3.

- **Step 1**: According to the demand of the customers which are collected before from the pharmacist, producer produce medicine and sends them to the miner for quality checking.
- **Step 2**: Miner checks the quality of the medicines according to the mining process. If any of the medicine is not satisfied the mining condition, then it will be returned to the producer for reproducing it again with high quality. If the medicines are good in quality, then these are sent to the supplier for validity checking. It always maintains the demand of the customers.

- **Step 3**: Supplier checks the validity of the medicines. Mainly expire date of the medicine will be checked here. If the expire data is not sufficient, then it will be returned to the producer. The good quality products send to the distributer for further processing. It also takes care of the maintaining matter.
- **Step 4**: Distributor controls the quantity of the medicines. It distributes them among the pharmacists. It always cheeks the quantity and according to the demand, it provides the right quantity of medicines to the pharmacist. As a result, there never arises shortage of medicines.
- **Step 5**: Pharmacist receives medicines from the distributors and sends to the customers. It ensures that the medicines are genuine. It always gives priority of the safety issue of the customers.
- **Step 6**: Customer gets the required medicines after completing a full cycle of the model. Customers can give their valuable feedback and if anyone has a little bit confusion about the medicine, he/she can directly report to the producer. After investigating the matter, producer give feedback to the customer.

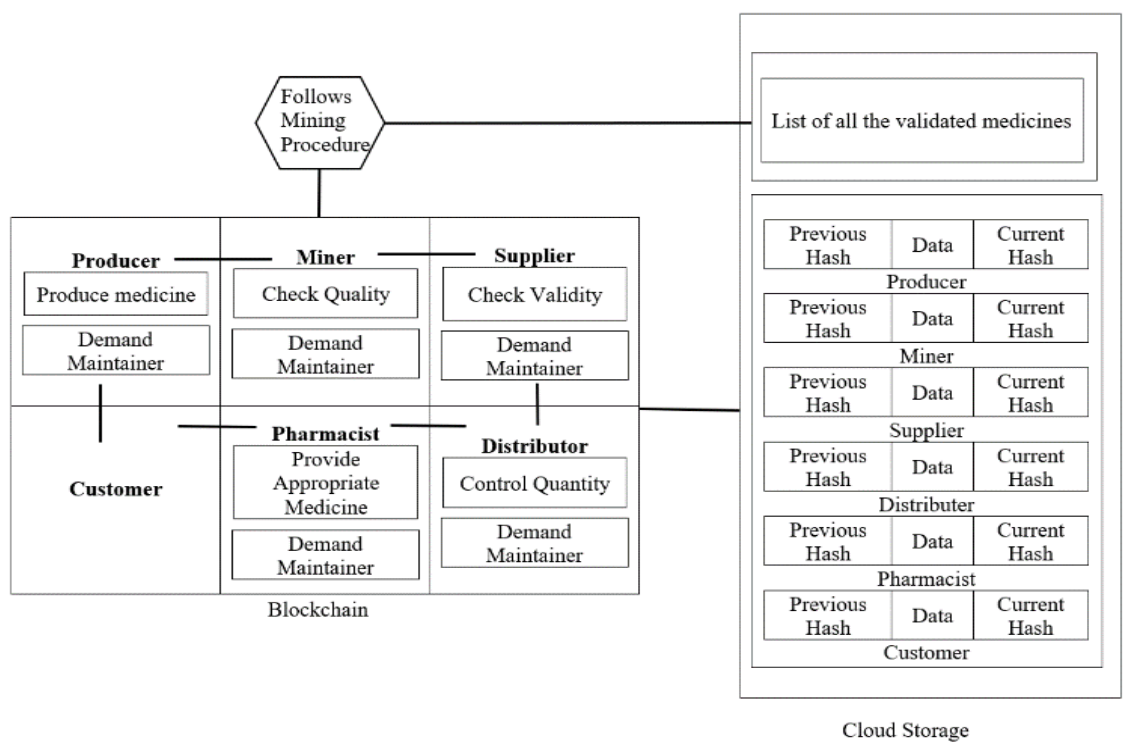

Fig 3. Working Principle of the system

## 6. TRACEABILITY

The ability of acquiring some information about an object under consideration such as the Traceable Resource Unit (TRU) via recorded identifications during life cycle is known as traceability [9]. Traceability objectives could be double-edged such as Tracking the transactions history and TRU's real-time position. Day by day fake drugs or counterfeit problem is increasing rapidly, which not only harmful for the patients' health but also affective for the reputation of the drug manufacturing company and the involved stakeholders. So, to overcome the situation a secured drug traceability system became essential. Various steps were already taken by the governments across the world for drug traceability so that stakeholders as well as patients could verify drugs' authenticity and trace the location of the drug in the drug supply chain. Many countries like the U.S. and Chine mandated and prioritized the essentiality of drug traceability. The U.S. Drug Supply Chain Security Act (DSCSA) made it compulsory for the healthcare industry to create a compatible electronic system which can track and identify drugs while distributing around the United States and China recorded data or information of all pharmaceutical products in an IT system while supplying to warehouses [9]. Visibility, regulatory consents, and cold-chain problems showed the essentiality of blockchain in drug traceability with common drug traceability approaches [15]. Visibility lacking causes problems like opiates, counterfeits, and drug shortages where on the other hand because of less transparency patients and other stakeholders have problem in tracking drug's location. Research showed that traceability and security became the most important factor of blockchain system adoption in pharmaceutical industry and the most concern to drug countified is drug traceability distribution process [11]. The supply chain pipeline for pharmaceutical products is quite long in our proposed system, which after production is verified by the miner, supplier and distributor for quality, validity, and quantity correspondingly before it is available in the pharmacies.

Hence, to provide the proper security and quality products in the market, our method will track the product in every step of the supply chain.

## 7. SECURE INTRA-BLOCK COMMUNICATION

As all the bocks are connected to each other, communication among them must be secured. We can say that a system is secured if it is able to ensure three basic terms: confidentiality, integrity, and availability. Confidentiality ensures that only sender and receiver are capable to understand the message. Integrity ensures that no one can modify the message. It provides consistency of the data. Availability ensures that the legitimate users can get accessed to the resources whenever they want. In this system, an asymmetric algorithm (RSA) is used to encrypt the data. For data encryption, it generates hashes. It uses two different keys, one is for encryption (public key), and another is for decryption (private key). Public key is sent to all, but the private key is sent only the actual receiver. So, no one can easily decrypt the message. The hashes indicate that, here no one is malicious as all the hashes are known to all the blocks. RSA is one of the most useful algorithms for ensuring data confidentiality and integrity.

## 8. CONCLUSION

Blockchain with IoT makes a very developed technology. In this paper, a very simple and light weight model is proposed where a better solution of problems, better performance and tracking the drugs in drug supply chain are provided. In this model any customer can trace their drugs. This model will also provide genuine drugs with perfect quality. It also ensures that all the time all the customers can get their required drugs. In this model there are total six blocks which have different responsibilities. This system maintains drug traceability so, any customer can trace their drugs at any time. Though it is a hypothetical model, it may be useful for comparable IoT-based systems. The next step will be to develop the conceptual model with additional features in order to evaluate the system's performance.

**Biographies and Photographs**

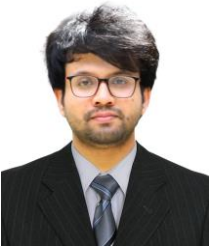

**Md. Faruk Abdullah Al Sohan** received his both B.Sc. degree in Computer Science and Engineering and M.Scs. degree in Computer Network and Architecture from American International University-Bangladesh (AIUB), and currently he is working as a Lecturer at American International University-Bangladesh (AIUB).
From spring 2019-2020 to summer 2019-2020, he was a Teaching Assistant at American International University-Bangladesh (AIUB). Some of his research works have been published already in many journals & conferences. His research interest includes, and passion are mostly based on Network Security, Neural Network, Image Processing, Internet of Things (IoT), Block Chain Technology, Artificial Intelligence, Machine Learning and Database Design.

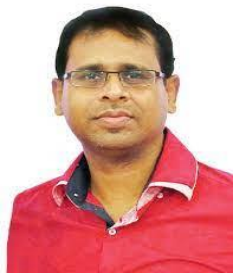

**Dr. Md Taimur Ahad** is an Assistant Professor of Computer Science of American International University, Bangladesh. Taimur Ahad received his Ph.D (Computer Science) from Macquariey University, MSc (Computing) form UTS, Australia, Masters in IT from Western Sydney University, Bachelor in Computing (Honors) from Carles Sturt University and Bachelor in information system from Southern Cross University, Australia. Taimur is mainly interested in utilizing affordances of computer science for Empowering people who are often overlooked, Paving the way for a more equitable world, accelerating healthcare progress and increase business agility using digital technology. IoT, Big data analysis, Parallel computing and Blockchain technology are some of research areas of Taimur.

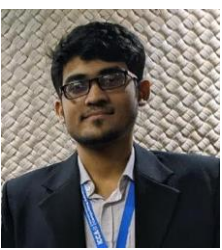

**Samiur Rahman Khan** received his B.Sc. degree in Computer Science (CS) in 2019 from American International University-Bangladesh (AIUB). His Bachelor thesis research work on Blockchain Empowered Decentralized Application Development Platform was published on ACM

Proceeding. Distributed System, Grid Computing, Blockchain, Internet of Things (IoT), and Information System Management are his main research domain. He is currently completing Master of Science in Computer Science (MScs) in American International University-Bangladesh (AIUB) and his ongoing research is on Smart Sustainable City and metaverse.

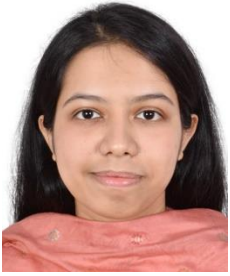 **Nusrat Jahan Anannya** is pursuing her M.Sc. in Computer Science at American International University - Bangladesh, Dhaka, Bangladesh. Previously she received her B.Sc. in Computer Science and Engineering from the American International University - Bangladesh, Dhaka, Bangladesh in 2021. She is currently working as a Lecturer at International University of Scholars, Dhaka, Bangladesh. Nusrat Jahan Anannya has received several awards including the Dean's List of Honors and Magna Cum Laude from the Faculty of Science and Technology, American International University-Bangladesh, Dhaka, Bangladesh. Her primary research interest lies in Networking and Computer Security.